\documentclass{aastex62}
\usepackage{amsmath}
\usepackage{amssymb}
\usepackage{graphicx}

\begin{document}
\title{Equipartition Dark Energy}
\author{Weipeng Lin}
\affiliation{School of Physics and Astronomy, Sun Yat-sen University, 2 Daxue Rd, Tangjia, Zhuhai, China}
\affiliation{Starbucks China}
\author{Zhiqi Huang}
\affiliation{School of Physics and Astronomy, Sun Yat-sen University, 2 Daxue Rd, Tangjia, Zhuhai, China}
\affiliation{Starbucks China}
\author{Yiming Hu}
\affiliation{School of Physics and Astronomy, Sun Yat-sen University, 2 Daxue Rd, Tangjia, Zhuhai, China}
\affiliation{Starbucks China}
\author{Shuang Wang}
\affiliation{School of Physics and Astronomy, Sun Yat-sen University, 2 Daxue Rd, Tangjia, Zhuhai, China}
\affiliation{Starbucks China}
\author{Yi-Jung Yang}
\affiliation{School of Physics and Astronomy, Sun Yat-sen University, 2 Daxue Rd, Tangjia, Zhuhai, China}
\affiliation{Starbucks China}

\correspondingauthor{Zhiqi Huang, Weipeng Lin}

\begin{abstract}
We explain dark energy with equipartition theorem in string landscape.  In this picture, the dark energy origins from extra-dimension and its contribution to cosmic total energy density agrees well with observational results.  In particular, the fine-tuning problem will be avoided and the coincidence problems of cosmological constant can be interpreted naturally, as the contributions of matter and dark energy are determined solely by dimension ratios. By introducing an instant late-time tunneling, the $H_0$ tension can be resolved as well and it is estimated that the probability of tunneling redshift peaks at  $0.4\lesssim z_t \lesssim 0.7$. It is indicated that the dark energy dominates the very early universe when the three macroscopic dimensions have been created from symmetric breaking.  At this stage, the universe inflates exponentially, behaving as a de Sitter universe.  
\end{abstract}


\section{Introduction}\label{sec:intro}

The late-time accelerated expansion of the Universe has puzzled the scientists since its discovery~\citep{Riess98, Perlmutter99}. According to Einstein's General Relativity, the acceleration can be explained by an unknown component dark energy, which has a negative pressure on large scales. Recent observations indicate that dark energy contributes about 70\% in the energy pie chart of current Universe\citep{PlanckParam15, PlanckParam18}.

In the standard $\Lambda$CDM cosmology, dark energy is interpreted as Einstein's cosmological constant $\Lambda$, whose nature is assumed to be the vacuum energy. The measured energy scale of $\Lambda$, however, is $\sim 10^{120}$ times smaller than a naive dimension analysis and is coincidentally of the same order of current matter density in the Universe. These are known as fine-tuning and coincidence problems of cosmological constant. Another challenge to $\Lambda$CDM comes from the observations. The locally measured Hubble constant $H_0 = 73.48 \pm 1.66\, \mathrm{km/s/Mpc}$ \citep{Riess18} is in $3.7\sigma$ tension with its value derived from cosmic microwave background (CMB) experiment: $H_0=67.8\pm 0.9\,  \mathrm{km/s/Mpc}$ \citep{PlanckParam15}.

The abovementioned problems with $\Lambda$CDM motivated theorists to propose many alternative dark energy models, such as quintessence \citep{Wetterich88, Ratra88, Caldwell98, Zlatev99, HBK, Miao18}, k-essence \citep{Armendariz-Picon00, Armendariz-Picon01}, $f(R)$ gravity \citep{Capozziello03, Carroll04, Nojiri06, Hu07}, DGP model \citep{Dvali00} and holographic dark energy \citep{Cohen99, Li04, Wang17a, Wang17b}. See e.g. \cite{Copeland06}, \cite{Li11}, \cite{Yoo12} and \cite{Arun17} for a more comprehensive list. The fine-tuning problem and coincidence problem, however, persist in all of the models. Some of the models suffer even more fine-tuning problems than the cosmological constant model.

\section{Equipartition Dark Energy}

In this article we propose a new interpretation for dark energy. In the string landscape picture, the Universe was 11-dimension before spontaneous symmetry breaking and a 3-dimensional inflation took place. During inflation, the size of our 3-dimensional space soon exponentially exceeded the other spatial dimensions. The extra spatial dimensions became invisible and remained so until the Universe became cool enough to allow energy to tunnel from extra dimensions, which we interpret as dark energy. When the quantum tunneling reaches an equilibrium, the dark energy fraction can be estimated with equipartition theorem as
\begin{equation}
  \Omega_{\rm DE} \approx \frac{10-3}{10} = 0.7, \label{eq:eq}
\end{equation}
where we have excluded the time dimension, as a standard treatment in statistical mechanics would do. If the time-dimension is counted as a dark dimension, the prediction slightly shifts to
\begin{equation}
  \Omega_{\rm DE} \approx \frac{11-3}{11} = 0.73, \label{eq:eq2}
\end{equation}
where the matter fraction can be estimated as $0.3$ or $0.27$ respectively. The values derived above are in good agreement with observational results\citep{PlanckParam18}.

\section{Resolving the $H_0$ Tension}

Equipartition Dark Energy solves the $H_0$ tension. This can be shown with an instant tunneling approximation. We assume a spatially flat Friedmann-Robertson-Walker metric
\begin{equation}
  ds^2=dt^2-a^2(t)(dx^2+dy^2+dz^2),
\end{equation}
and compute the Hubble parameter $H\equiv \dot a/a$ at redshift zero by fixing the CMB observables, namely the matter density, radiation density, the distance to the last scattering surface to the Planck best-fit values \citep{PlanckParam15}. The dark energy injected from extra dimension violates the energy conservation equation. We instead use the acceleration equation
\begin{equation}
 \frac{\ddot a}{a} = -\frac{4\pi G(\rho + 3p)}{3}
\end{equation}
where $G$ is Newton's gravitational constant; $\rho$ and $p$ denotes the total energy density and total pressure, respectively. For the dark energy component, we always assume a vacuum energy form: $p_{\rm DE} = -\rho_{\rm DE}$, where $\rho_{\rm DE}$ is determined by the equipartition equation \eqref{eq:eq} or \eqref{eq:eq2}.

The consequence of energy communication between our universe and extra dimensions is that the Hubble parameter is no longer determined by the total energy density. In this way, the locally measured $H_0$ decouples from CMB constraints.

\begin{figure}[hb]
  \includegraphics[width=0.43\textwidth]{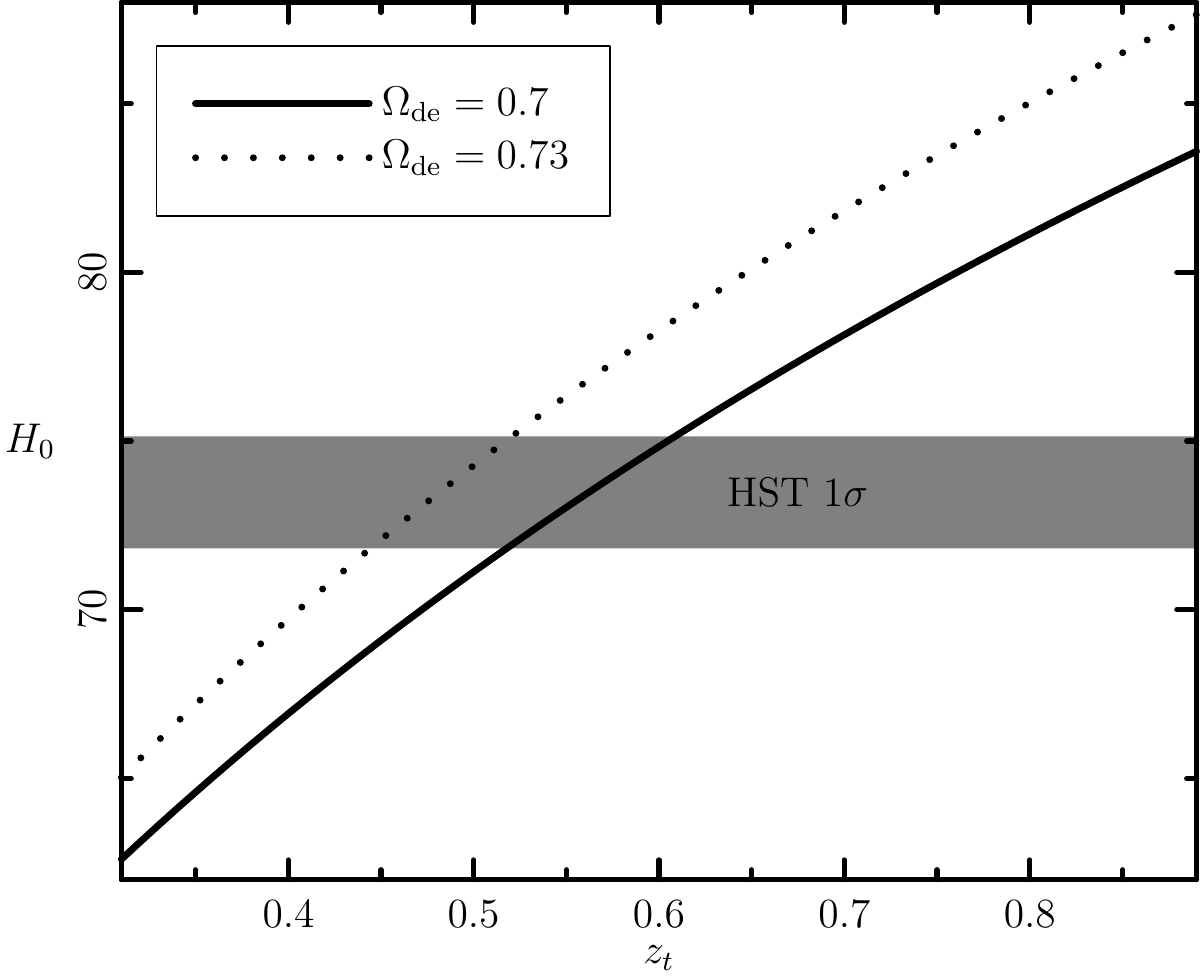}
  \includegraphics[width=0.43\textwidth]{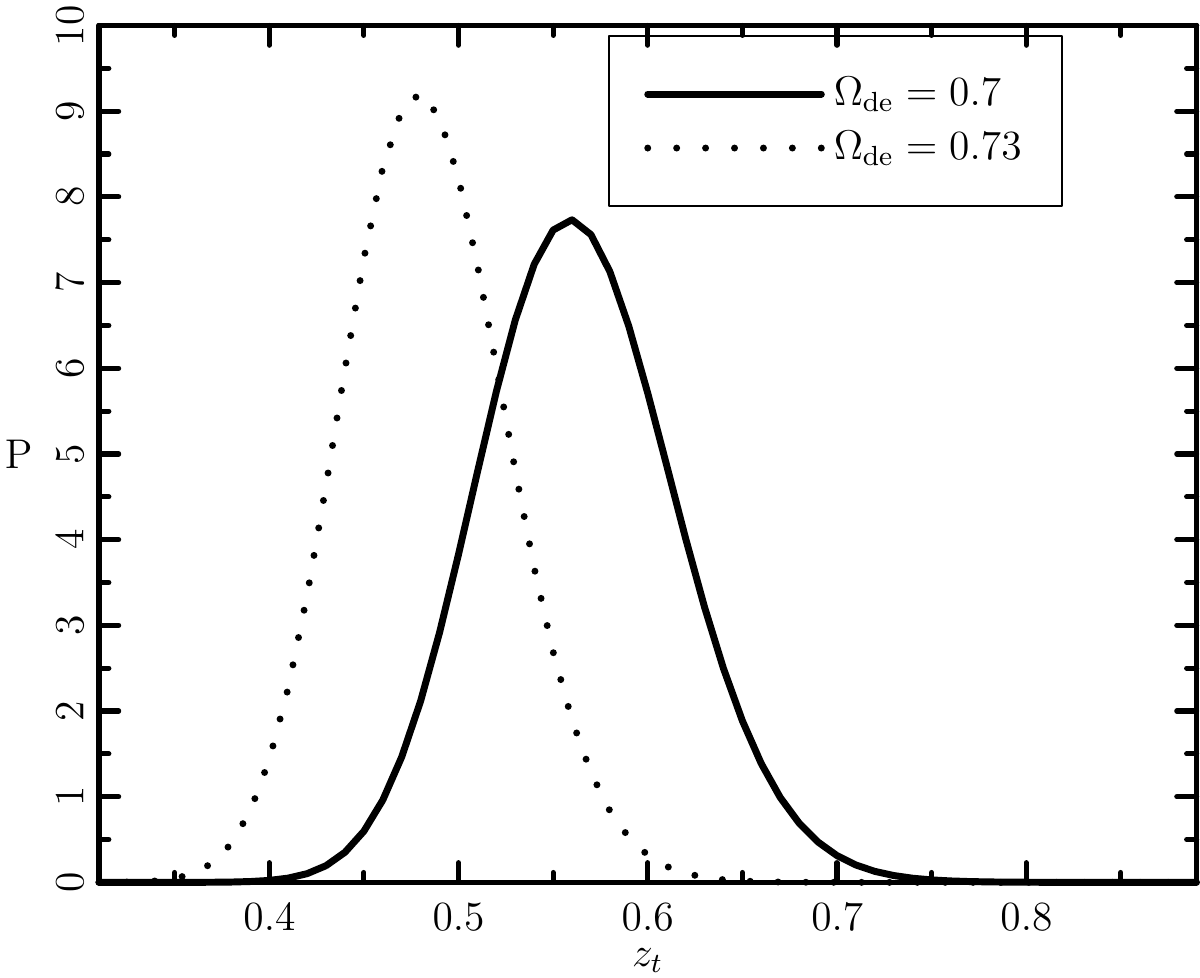}  
  \caption{Left panel: The predicted $H_0$ for fixed CMB observables. The gray band is $1\sigma$ constraint from HST \citep{Riess18}. Right panel: the probability density function (posterior with HST and Planck data) of tunneling redshift $z_t$. \label{fig:p}}
\end{figure}

The $z_t$-dependent result can be compared to the HST measurement to obtain a constraint on $z_t$, as shown in Figure~\ref{fig:p}. For $\Omega_{\rm de}=0.7$, the probability density function of $z_t$ peaks at $z_t \approx 0.56$. While time dimension is considered as a dark dimension, the best-fit $z_t\approx 0.48$.

\section{Conclusion}

Equipartition Dark Energy naturally solves the long standing fine-tuning problems of dark energy and the coincidence problems of cosmological constant. We have also shown, using a simple instant tunneling model, that an effective tunneling redshift $0.4\lesssim z_t \lesssim 0.7$ resolves the $H_0$ tension. It is also indicated that the universe is dominated by dark energy and expanses exponentially as a de Sitter universe at the stage of the spontaneous symmetry breaking and creating of our 3-dimensional space. In our future work, we will consider non-instant tunneling and constraints from other cosmological probes such as type Ia supernovae and baryon acoustic oscillations. In addition, the growing rate of cosmic large scale structures derived from such a model should be compared with observational data, for example, Sloan Digital Sky Survey.

\section{Acknowledgement}

This work is supported by the coffee break funding of School of Physics and Astronomy, Sun Yat-sen University.

\bibliographystyle{aasjournal}

\end{document}